\documentclass[aps,prl,showpacs,amsmath,preprint,superscriptaddress,floatfix,nobalancelastpage,footinbib]{revtex4-2}
\usepackage[colorlinks=true, citecolor=black, linkcolor=black, urlcolor=black]{hyperref}
\usepackage[all]{hypcap}
\usepackage{amsmath}
\usepackage{enumerate,bm}
\usepackage{graphicx}
\usepackage{color}
\usepackage{esint}
\usepackage[usenames,dvipsnames]{xcolor}
\usepackage{subfigure}
\usepackage{flushend}
\usepackage{amssymb}
\usepackage{float}
\usepackage{dcolumn}% Align table columns on decimal point
\usepackage{ulem} %striketrhough

\usepackage{setspace}
\usepackage{verbatim}

\begin{document}
%\linenumbers

\author{Dongbin~Shin}
\email{dshin@gist.ac.kr}
\affiliation{Max Planck Institute for the Structure and Dynamics of Matter and Center for Free Electron Laser Science, 22761 Hamburg, Germany}
\affiliation{Department of Physics and Photon Science, Gwangju Institute of Science and Technology (GIST), Gwangju 61005, Republic of Korea}

\author{Simone~Latini}
\affiliation{Max Planck Institute for the Structure and Dynamics of Matter and Center for Free Electron Laser Science, 22761 Hamburg, Germany}

\author{Christian~Sch\"afer}
\affiliation{Department of Microtechnology and Nanoscience, MC2, Chalmers University of Technology, 412 96 G\"oteborg, Sweden}

\author{Shunsuke~A.~Sato}
\affiliation{Max Planck Institute for the Structure and Dynamics of Matter and Center for Free Electron Laser Science, 22761 Hamburg, Germany}
\affiliation %[CCS]
{Center for Computational Sciences, University of Tsukuba, Tsukuba 305-8577, Japan}

\author{Edoardo~Baldini}
\affiliation{Department of Physics, University of Texas at Austin, Austin, TX, USA}

\author{Umberto~De~Giovannini}
\affiliation{Max Planck Institute for the Structure and Dynamics of Matter and Center for Free Electron Laser Science, 22761 Hamburg, Germany}
\affiliation{Universit\`a degli Studi di Palermo, Dipartimento di Fisica e Chimica—Emilio Segr\`e, via Archirafi 36, I-90123 Palermo, Italy}

\author{Hannes~H\"ubener}
\affiliation{Max Planck Institute for the Structure and Dynamics of Matter and Center for Free Electron Laser Science, 22761 Hamburg, Germany}

\author{Angel~Rubio}
\email{angel.rubio@mpsd.mpg.de}
\affiliation{Max Planck Institute for the Structure and Dynamics of Matter and Center for Free Electron Laser Science, 22761 Hamburg, Germany}
\affiliation{Nano-Bio Spectroscopy Group, Departamento de Fisica de Materiales, Universidad del País Vasco UPV/EHU- 20018 San Sebastián, Spain}
\affiliation{Center for Computational Quantum Physics (CCQ), The Flatiron Institute, 162 Fifth avenue, New York NY 10010.}

\title{Simulating terahertz field-induced ferroelectricity in quantum paraelectric SrTiO$_3$}

%%%.  3473 (Maximum length is 3750)%% now it is fine !!02/01/2020 21:33%%%%%

\date{\today}

\begin{abstract} 
Recent experiments have demonstrated that light can induce a transition from the quantum paraelectric to the ferroelectric phase of SrTiO$_3$.
Here, we investigate this terahertz field-induced ferroelectric phase transition by solving the time-dependent lattice Schr\"odinger equation based on first-principles calculations.
We find that ferroelectricity originates from a light-induced mixing between ground and first excited lattice states in the quantum paraelectric phase.
In agreement with the experimental findings, our study shows that the non-oscillatory second harmonic generation signal can be evidence of ferroelectricity in SrTiO$_3$.
We reveal the microscopic details of this exotic phase transition and highlight that this phenomenon is a unique behavior of the quantum paraelectric phase.
\end{abstract}

\maketitle

Ultrafast light-induced dynamics provide attractive physical phenomena beyond the steady-state of materials. 
Recent studies have shown that driving materials out of their original equilibrium state can induce complex phase transitions or largely modify properties and functionalities~\cite{Mitrano2016,Nova2017,Li2019,Nova2019,Kogar2020,Jin2019}.
Theoretical works have suggested that a large number of terahertz (THz) field-induced phenomena are closely related to nonlinear phonon interactions~\cite{Knap2016,Kennes2017,Guan2021,Shin2020}.
For example, by applying a THz pulse, a high temperature superconducting state has been experimentally observed in K$_3$C$_{60}$ which otherwise exhibits a low critical temperature in equilibrium conditions~\cite{Mitrano2016,Cantaluppi2018,Budden2021}.
Possible topological phase transitions induced by THz pulses have also been demonstrated in layered materials such as $T_d$-WTe$_2$, ZrTe$_5$, and graphene~\cite{Sie2019,Vaswani2020,McIver2020}.
The THz field-induced ferroelectric switching and structural transition through nonlinear phonon interactions have been investigated with a semi-classical approach~\cite{Subedi2014,Subedi2015,Juraschek2018,Mankowsky2017,Nova2019,abalmasov2020,Zhou2020}.
Another paradigmatic manifestation of the THz-matter interaction is the transformation of paraelectric SrTiO$_3$ into its ferroelectric phase~\cite{Li2019,Nova2019}.

Unlike other transition metal perovskite materials such as BaTiO$_3$ and PbTiO$_3$, at low-temperature SrTiO$_3$ does not show a paraelectric-to-ferroelectric phase transition~\cite{Shirane1952,Samara1971,Bellaiche2000,Guo2000,Miyasaka2003,Haeni2004,Fennie2006,Lee2010,Muller:1979je,Zhong1995,Shin2021}.
The origin of this phenomenon lies in the potential energy surface of SrTiO$_3$ which along the ferroelectric soft (FES) mode is characterized by a shallow double-well potential at low temperature ~\cite{Muller:1979je,Zhong1995,Shin2021}.
Despite classically favoring a ferroelectric phase, SrTiO$_3$ remains in the high symmetry phase due to nuclear quantum fluctuations~\cite{Zhong1995,Shin2021}.
The corresponding low-temperature phase of SrTiO$_3$ is therefore a quantum paraelectric phase.
Recent first-principles calculations revealed that quantum fluctuations and the nonlinear interaction between the FES mode and lattice strain are required to describe the quantum paraelectric phase in SrTiO$_3$~\cite{Shin2021}.
With these effects, the computational treatment is capable of reproducing temperature-dependent properties, such as the energy of the FES mode and the dielectric constant~\cite{Shin2021,Muller:1979je}.
Relying on such a first-principles description, an optically stabilized ferroelectric ground state of SrTiO$_3$ in a cavity has been theoretically demonstrated~\cite{Simone2021}.

Here, we investigate the THz field-induced non-equilibrium ferroelectricity of the quantum paraelectric state of SrTiO$_3$ based on a lattice model derived from extensive first-principles calculations~\cite{Shin2021}. 
Solving the time-dependent lattice Schr\"odinger-Langevin equation shows that a ferroelectric state can be obtained in SrTiO$_3$ by illumination with a single cycle THz pulse, as shown in Fig.~1(a).
We find that this ferroelectricity originates from the mixing between ground and first excited states of the lattice wavefunction.
Based on time-dependent density functional theory (TDDFT) calculations, we obtain a non-oscillatory second harmonic generation (SHG) signal and find that it is a signature of the induced broken symmetry, in agreement with the conclusions drawn from experimental observations~\cite{Li2019,Nova2019}.
Lastly, we discuss the conditions of THz pulse frequency and dissipation rate for ferroelectricity.

\begin{figure}[t!]
  \centering
  {\includegraphics[width=0.6\textwidth]{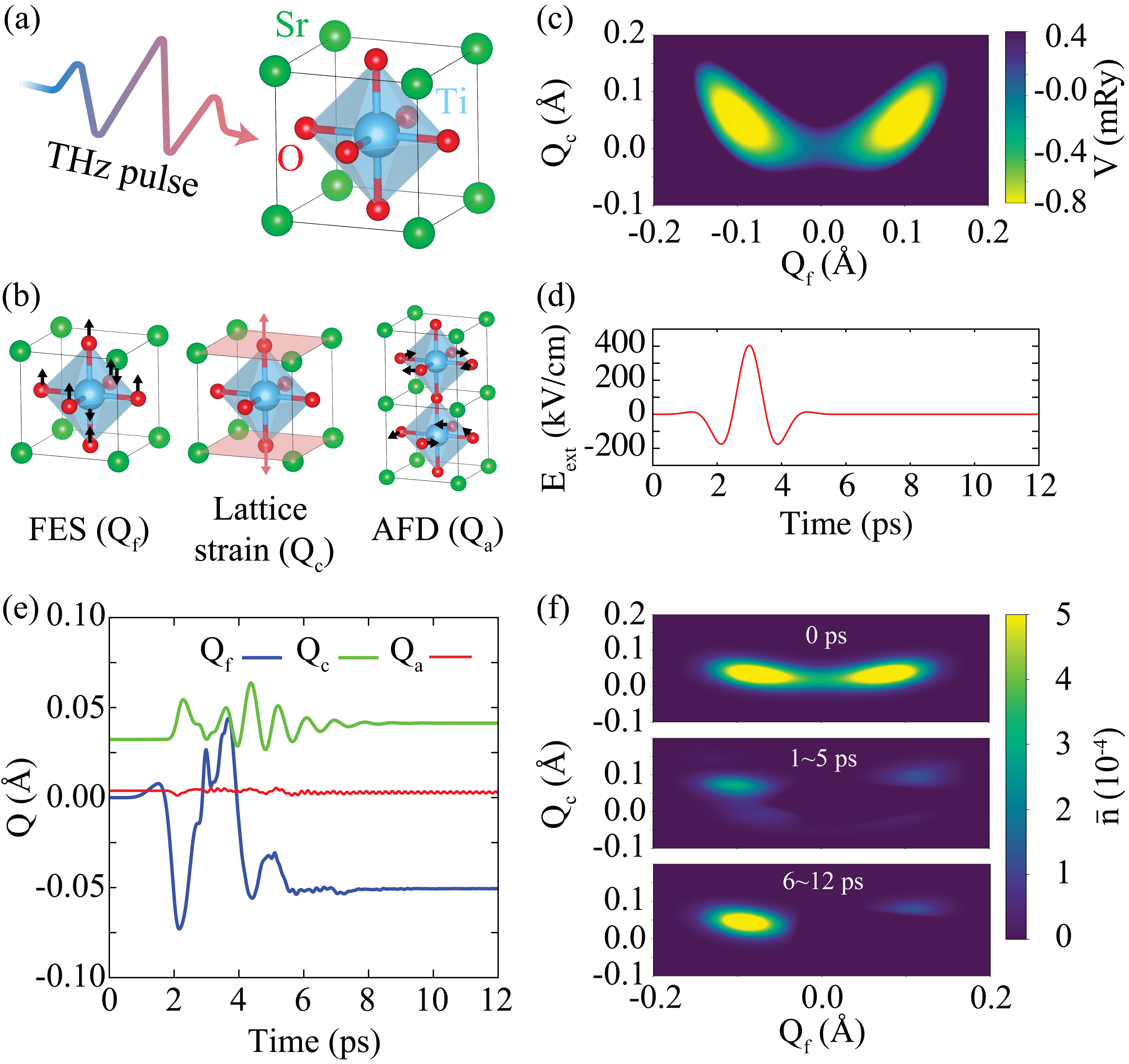}}
   \caption{
   THz field-induced ferroelectricity in SrTiO$_3$.
   (a) Schematic image of the THz pulse applied to SrTiO$_3$.
   (b) Schematic image of the atomic displacements involved in the FES mode ($Q_f$), lattice strain ($Q_c$), and AFD mode ($Q_a$).
   (c) Two-dimensional potential energy surface of SrTiO$_3$ along with the FES mode and lattice coordinates.
   (d) Time profile of single-cycle THz pulse with $0.5$~THz frequency.
   (e) Time profiles of the expectation value of the FES, AFD modes, and c-axis perturbed by the THz pulse.
   (f) Variation of the time-averaged lattice density perturbed by the THz pulse.
}
\end{figure}

To investigate the THz field-induced ferroelectric transition from quantum paraelectric SrTiO$_3$, we first construct the potential energy surface for a relevant set of SrTiO$_3$ phonon modes by means of density functional theory calculations.
In our recent first principles-based study, we reported that the quantum paraelectric state in SrTiO$_3$ can be described by a two-dimensional (2D) lattice Schr\"odinger equation~\cite{Shin2021}.
The related 2D effective potential, which describe the FES mode ($Q_f$) and the lattice expansion/contraction along the FES mode direction ($Q_c$) as shown in Figs.~1(b) and~1(c), provides a description of the FES mode in quantum paraelectric SrTiO$_3$ which  is consistent with experimental observations~\cite{Muller:1979je,Vogt1995,Yamanaka2000}.
The effective potential energy surface is constructed from fitting the calculated total energy as a function of geometry distortions as follows: $\hat{V}^{\text{f,c}}=\sum_{i=1}^{12} k_{f,i}\hat{Q}^{2i}_f + \sum_{j=2}^{10} k_{c,j}\hat{Q}^{j}_c+\sum_{i=1}^{12} \sum_{j=1}^{10} k_{fc,i,j}\hat{Q}^{2i}_f \hat{Q}^{j}_c$~\cite{Shin2021,SM}.
%~\cite{Sophya2013,Chou2015,Riseborough1985,Shi2020,Mollow1970,Kozina2019,Ghosez1998,Shah2008,Souza2003,Yabana2011,Gogoi2016,Guo2011,Tancogne2020,Shirane1969,Aschauer:2014,Wojdel2013}
In addition, we include the dynamical effect of the antiferro-distortive (AFD) mode, which shows a strong nonlinear phonon interaction with the FES mode~\cite{Aschauer:2014,Li2019}, by performing additional Verlet integrations according to: $M_a\ddot{Q}_{a}=-dV_{a}/dQ_{a}$ with $V_{a}=\sum_{i=1}^{10} k_{a,i}Q_{a}^i +\sum_{i=1}^{10}\sum_{j=1}^{12}k_{af,i,j}Q_{a}^i\langle Q_{f}\rangle ^{2j}+\sum_{i=1}^{10}\sum_{j=1}^{10}k_{ac,i,j}Q_{a}^i\langle Q_{c} \rangle ^{j}$ and effective mass for AFD mode ($M_a=2.12 \times 10^{-25}$~kg).
The displacement of the AFD mode ($Q_a$) is obtained at given time $t$ for the expectation values of the FES mode ($\langle Q_f \rangle $) and c-axis expansion/contraction ($\langle Q_c \rangle$). 
This solution is then used to solve the lattice Sch\"odinger-Langevin equation with the additional time-dependent potential $\hat{V}^{f,c,a}[\hat{Q}_f,\hat{Q}_c,t]=\sum_{i=1}^{10}\sum_{j=1}^{12}k_{af,i,j}Q_{a}^i(t)\hat{Q}_{f}^{2j}+\sum_{i=1}^{10}\sum_{j=1}^{10}k_{ac,i,j}Q_{a}^i(t)\hat{Q_{c}}^{j}$.
We evaluate all DFT total energies with the QUANTUM ESPRESSO package with the projector augmented wave method and a plane-wave basis set with 70 Ry energy cut-off~\cite{Giannozzi2017}.
The electron-electron exchange and correlation potentials are described through the Perdew-Berke-Ernzerhof functional~\cite{Perdew1996}.
We consider the $\sqrt{2} \times \sqrt{2} \times 2$ tetragonal unit cell for the low temperature phase of SrTiO$_3$ and we sample the Brillouin zone with a $6 \times 6 \times 4$ $\mathbf{k}$-point grid.
Details on the parameters used for the 2D effective potential energy surface are summarized in the Supplemental Material~\cite{SM}.

To describe the effect of the THz pump pulse on quantum paraelectric SrTiO$_3$, we solve the time-dependent 2D lattice Schr\"odinger-Langevin equation $i\hbar \frac{d}{dt}\psi=\hat{H}_{\text{2D}}\psi+\gamma(\hat{S}-<\hat{S}>)\psi$, with the Hamiltonian $\hat{H}_{\text{2D}}[\hat{Q}_f,\hat{Q}_c,Q_a(t),t]={\hat{P}^2_f}/{2M_f}+{\hat{P}^2_c}/{2M_c}+\hat{V}^{\text{f,c}}[\hat{Q}_f,\hat{Q}_c]+\hat{V}^{\text{f,c,a}}[\hat{Q}_f,\hat{Q}_c,Q_a(t)]+E_{ext}(t)Z_f\hat{Q}_f$. 
In parallel, we evaluate the Newton equation of motion for the AFD mode, and update the potential ($\hat{V}^{\text{f,c,a}}[\hat{Q}_f,\hat{Q}_c,Q_a(t)]$) with updated $Q_a$ at $t$.
The effective mass of the FES mode is $M_f=1.76 \times 10^{-25}$~kg and the mass of the tetragonal cell is used as the lattice effective mass ($M_{c}=\sum_i M_i=1.22\times 10^{-24}$~kg)~\cite{Shin2021}.
The $\gamma$ and $\hat{S}[Q_f,Q_c]=arg[\psi(Q_f,Q_c,t)]$ are the dissipation rate and the phase of the wavefunction, respectively~\cite{Sophya2013,Chou2015}.
In our simulation, the motion of the FES mode, the out-plane lattice, and AFD mode are taken into account explicitly, while the phonon-phonon scattering effects of all the other modes~\cite{Kozina2019,Li2019,Juraschek2018} are described by the fractional dissipation Langevin term. 
Notably, a fractional Langevin equation has been employed to describe the heat dissipation into harmonic and anharmonic oscillators such as an Ohmic bath~\cite{Riseborough1985,Shi2020,Mollow1970,Kozina2019}.
We use a dissipation rate of $\gamma=1.2$~THz for the FES mode evaluated from the $ab~initio$ molecular dynamics simulation~\cite{SM}, which is also in agreement with the value reported by recent spectroscopic measurements~\cite{Kozina2019}.
We consider the interaction between the FES mode and the THz pulse ($H_{int}=E_\text{eff}(t)Z_f\hat{Q}_f$) to be linear in the effective field strength and in the electric dipole generated by $Q_f$ with its mode effective charge ($Z_f=37.0e$).
The mode effective charge is evaluated from the Born effective charge and verified from the modification of the potential energy surface by applying the finite E-field in the periodic boundary condition~\cite{Souza2003}.
In the cubic unit cell, the value of mode effective charge in the normal coordinate is $Z_\text{cubic}=0.91~e~amu^{-1/2}$, consistently with the value in other perovskite oxides~\cite{Subedi2017}.
The effective field strength ($E_\text{eff}(t)=0.58 E_\text{ext}(t)$) in the medium is evaluated from the dielectric function of SrTiO$_3$~\cite{Yabana2011,Gogoi2016,Kozina2019,Guo2011}, when $E_\text{ext}$ is external field strength.
Details on the time-dependent Schr\"odinger-Langevin equation, the mode effective charge, and the effective field strength are discussed in the SM~\cite{SM}.

We simulate the action of a single-cycle THz pulse on the low-temperature quantum paraelectric state of SrTiO$_3$ by choosing it to be resonant with the frequency of the FES mode ($0.5$~THz) and featuring a maximum strength of $400$~kV/cm, as shown in Fig.~1(d).
This THz pulse perturbs the quantum paraelectric ground state $\psi_0(Q_f,Q_c,t=0)$ and induces a variation in the expectation values of $Q_f(t)$, $Q_c(t)$, and $Q_a(t)$, as shown in Fig.~1(e).
Notably, the asymmetric potential along $Q_c$ provides a non-zero expectation value of $\hat{Q}_c$ ($Q_c(t=0)=31$~m\AA) in the ground state because of quantum fluctuations~\cite{Shin2021}.
The time profile of the FES mode exhibits a non-oscillatory component after $6$~ps. 
Because the FES mode is directly related to the electric polarization of the system ($P=Z_f Q_f$), this plateau indicates that the single-cycle THz pulse induces a ferroelectric transition.
We also find that the THz pulse elongates the lattice ($Q_c$), providing a deeper double well for the FES mode, as shown in Fig.~1(c)~\cite{Shin2021}, while the displacement of AFD mode is negligible $|\delta Q_a| = 1$~m\AA$\sim0.03^{\circ}$.
To understand the variation of the 2D lattice state, we evaluate the time-averaged lattice densities ($\bar{n}(t_1,t_0)=\int_{t_0}^{t_1}|\psi(Q_f,Q_t,t)|^2 dt$) for a given period of time, as displayed in Fig.~1(f).
At the ground state ($t=0$~ps), the time-averaged lattice density is delocalized over the double-well potential along the $Q_f$ direction, indicating a quantum paraelectric ground state.
The applied THz pulse perturbs the lattice density at the time delay $1$~ps~$ < t < 5$~ps.
Upon application of the THz pulse ($t>6$~ps), the lattice density localizes in one of the two wells signaling the emergence of a ferroelectric state.

\begin{figure}[t!]
  \centering
  {\includegraphics[width=0.5\textwidth]{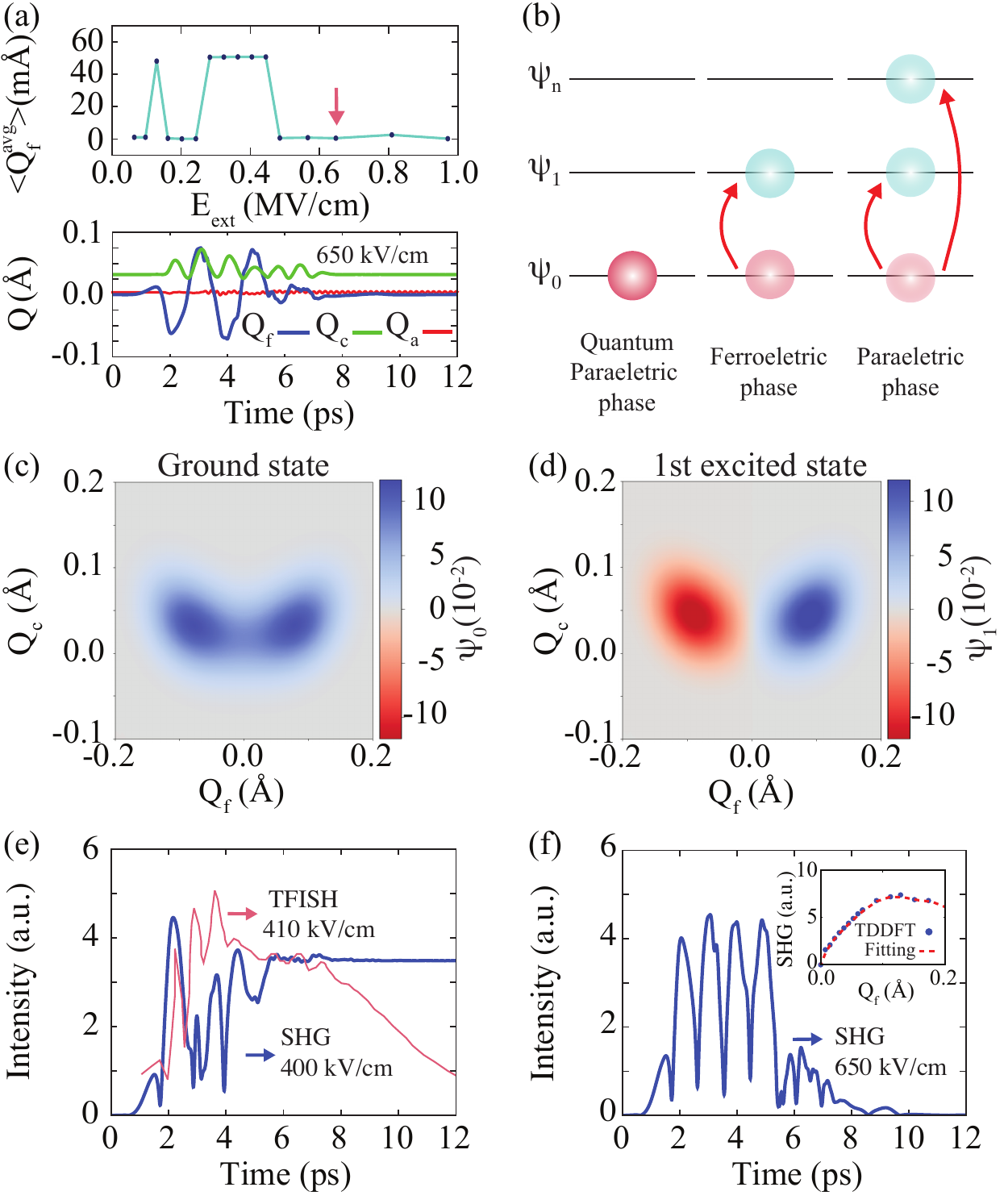}}
   \caption{
   Mechanism of the THz field-induced phase transition from the quantum paraelectric state.
(a) Time-averaged value of the FES displacement($Q_f^{avg}$) perturbed by a THz pulse in a wide range of intensities (top panel) and time-profile of $Q_f$, $Q_c$, and $Q_a$ with a $650$~kV/cm pulse (bottom panel).
   (b) Schematic image of the phase transition from the quantum paraelectric state.
   (c) Ground and (d) first excited states of the quantum paraelectric phase in SrTiO$_3$.
   Time profile of the SHG signal induced by a THz pulse with (e) $400$~kV/cm and (f) $650$~kV/cm.
  In (a), the pink arrow indicates the result with $650$~kV/cm pulse.
  In (b), the red arrows reveal the THz field pulse-induced excitations of lattice wavefunction.
  In (e), the data of THz field–induced second harmonic signal with field strength $400$~kV/cm is obtained from Ref.~\cite{Li2019} and their intensities are scaled for a qualitative comparison.
  The inset of (f) shows the relation between the SHG signal and the FES mode distortion evaluated by TDDFT calculations.}
\end{figure}

This THz field-induced ferroelectric transition originates from the excitation between the ground and the first excited states.
In order to quantify the THz field-induced ferroelectricity, we evaluate the time-averaged $Q_f$ value as follows: $Q_f^{avg}=\frac{1}{t_1}\int^{t_0+t_1}_{t_0}Q_f(t)dt$, where $t_0=6$~ps is the time at the end point of the pulse and $t_1=12$~ps. 
As shown in the upper panel of Fig.~2(a), the THz field pulse induces ferroelectricity in a wide intensity range $0.15$~MV/cm~$< E < 0.50$~MV/cm.
The THz field-induced ferroelectricity and its field strength dependence can be explained in terms of phonon excitations by the THz pulse, as displayed in Fig. 2(b).
The THz pulse with appropriate strength allows excitation to the first excited state ($\psi_1$) from the ground state ($\psi_0$) shown in Figs. 2(c) and 2(d).
It is worth noting that the linear combination between ground and first excited states provides two degenerate ferroelectric states, which have opposite electric polarization ($\psi_0+\psi_1$ and $\psi_0-\psi_1$).
When a single-cycle THz pulse is applied, it breaks the symmetry of the excited state along $Q_f$, and ferroelectricity emerges. 
A high field strength THz pulse promotes transitions to excited states with much higher energy. As a result, it produces a light-mixed state equivalent to a thermally excited paraelectric state, spreading over the full double-well potential.
In the SM, we discuss details about the ferroelectric lattice wavefunction in the quantum paraelectric phase, and the direction of electric polarization~\cite{SM}.

In recent experimental results~\cite{Li2019,Nova2019}, the THz field–induced second harmonic (TFISH) and SHG signals of a delayed near-infrared probe have been interpreted as a sign of light-induced ferroelectricity in SrTiO$_3$.
To evaluate the SHG signal from the lattice wavefunction, we calculate the relationship between the SHG signal and $Q_f$ from TDDFT calculations with distorted geometry along the FES mode direction (see the inset of Fig.~2(f))~\cite{Tancogne2020}.
With tetragonal SrTiO$_3$ geometries distorted along the FES mode direction, we evaluate the current response by applying a $1.55$~eV probe pulse with $40$~fs duration.
From the fitted polynomial function ($\rm{SHG}(Q_f)=\sum_i^3 c_i|Q_f|^i$), the time-profile SHG signal during the THz field dynamics is estimated and compared with the experimentally observed TFISH signal obtained from Ref.~\cite{Li2019}, as shown in Fig.~2(e).
The simulated SHG signal induced by the $400$~kV/cm THz pulse exhibits the non-oscillatory plateau behavior similar to the time profile of $Q_f$.
The experimentally observed TFISH signal shows the emergence of a non-oscillatory signal, but the signal starts decaying after $8$~ps.
This inconsistency originates from the lack of finite-temperature heat bath in our system and the over-simplified description of the phonon-phonon scattering process in our simulation.
These points are not able to capture the long-time decay into the quantum paraelectric ground state from the transient ferroelectric state as the experiment.
 It is expected that the time-dependent path-integral Monte Carlo simulation or Schr\"odinger equation with heat bath could provide more realistic THz field-induced dynamics that includes transient behavior as experiment~\cite{Zhang2020,Muhlbacher2008,Caldeira1983}.
While both observable from the SHG and TFISH measurements are sensitive to the absence of the inversion symmetry, the comparison cannot be quantitative. 
Nevertheless, we can find a qualitatively similar behavior between the two responses, indicating consistent time scales for the emergence of ferroelectricity before $8$~ps.
On the other hand, the higher THz pulse intensity ($650$~kV/cm, Fig. 2(f)) induces the highly excited FES mode that leads to a simple oscillatory and decay of the SHG signal.
Because the highly excited state consists of the higher excited states rather than the first excited state, it is hard to form the ferroelectric state and decays into the quantum paraelectric ground state.
Similar to the experimental observations~\cite{Li2019,Nova2019}, our TDDFT calculation reveals that the non-oscillatory behavior of the SHG signal can be a vital fingerprint of the THz field-induced ferroelectricity.
In the Supplemental Material, we discuss details on the procedure used to evaluate the SHG signal from TDDFT calculations~\cite{SM}.

\begin{figure}[t!]
  \centering
  {\includegraphics[width=0.6\textwidth]{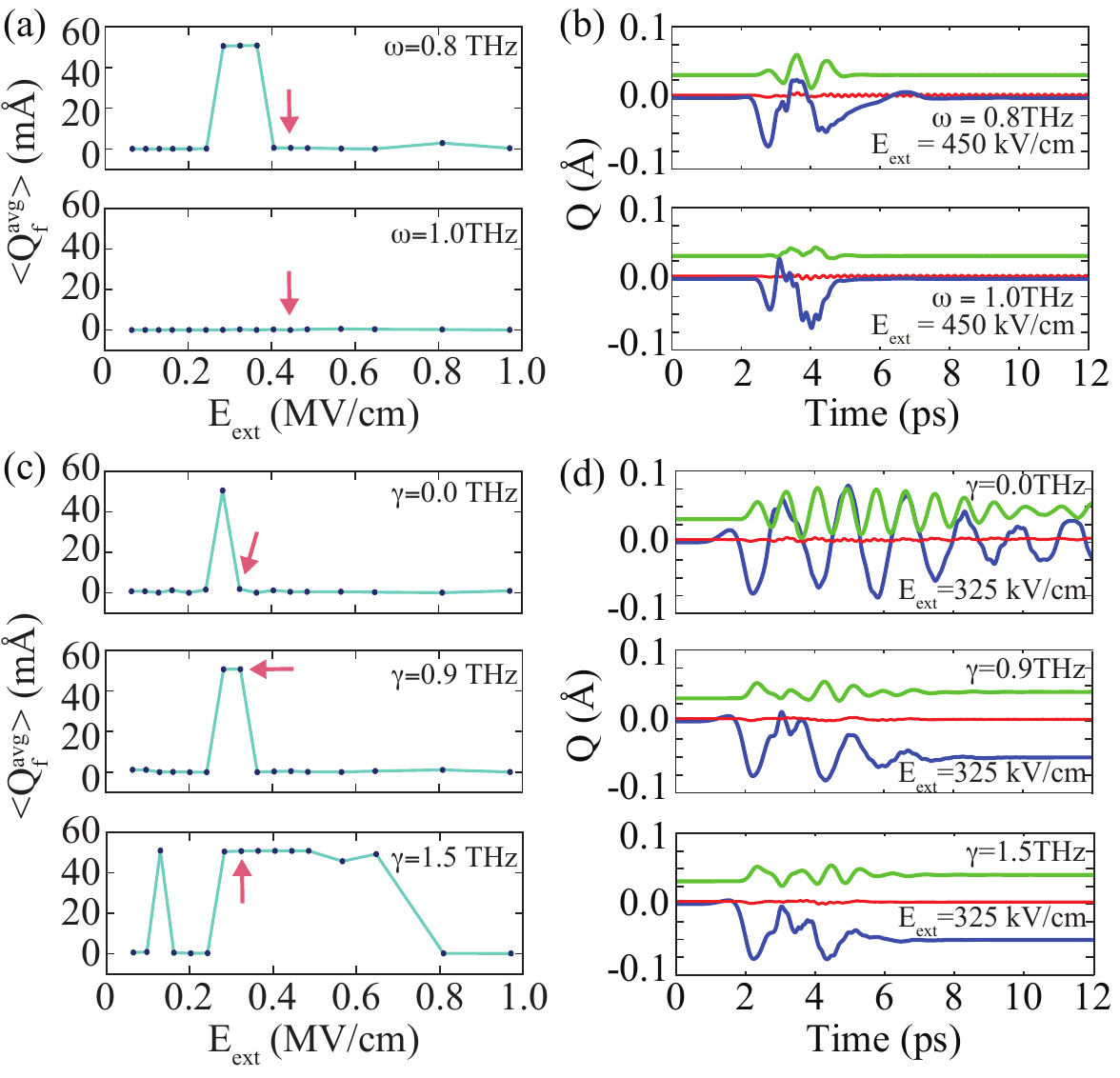}}
   \caption{
   Effects of frequency and dissipation on THz-induced dynamics in SrTiO$_3$.
   (a) Time-averaged value of FES displacement and (b) time profiles of $Q_f$, $Q_c$, and $Q_a$ at various THz frequencies $\omega$ and $\gamma =1.2$~THz.
   (c) Time-averaged value of FES displacement and (d) time profiles of $Q_f$, $Q_c$, and $Q_a$ at various dissipation rate $\gamma$ and $\omega=0.5$~THz.
   In (a) and (c), the pink arrows indicate the specific points for the time-profiles in (b) and (d).
}
\end{figure}

We investigate the impact of THz field spectral coverage on ferroelectricity.
When we apply a resonant THz pulse ($\omega=0.5$~THz), we observe non zero $Q_f^{avg}$ values in a wide range of THz field strength, as shown in Fig.~2(a).
On the other hand, a near-resonant single-cycle THz pulse with $\omega=0.8$~THz provides a suppressed range for non zero $Q_f^{avg}$ values, and an off-resonant THz pulse with $\omega=1.0$~THz gives negligible $Q_f^{avg}$ values, as shown in Fig.~3(a).
As a consequence, while resonant $\omega=0.5$~THz condition leads to a non-oscillatory behavior of $Q_f(t)$ at given field strength ($450$~kV/cm), there is no plateau with off-resonant ($\omega=1.0$~THz) and near-resonant $\omega=0.8$~THz frequencies, as shown in Fig.~3(b).
This result reveals that the resonant THz pulse effectively excites the FES mode from the ground state to the first excited state, and a superposition of the ground and first excited states forms ferroelectricity rather than an incoherent lattice wave shift.

We also investigated the dependence of the dissipation rate $\gamma$ on the THz field-induced ferroelectricity, as shown in Fig.~3(c).
In our simulation, we include the dissipation effect into the AFD mode, the out-plane motion of the lattice, and express Langevin dissipation through the phenomenological parameter $\gamma$.
When changing the dissipation rate $\gamma$, we still observe THz-induced ferroelectricity, but in a different field strength range.
With a lower dissipation rate ($\gamma=0$ and $0.9$~THz), SrTiO$_3$ shows ferroelectricity within a smaller range of field strengths and simple oscillations of $Q_f$ and $Q_c$ (see Fig.~3(d)).
It originates from the oscillation of highly excited FES modes over the double well.
On the other hand, higher dissipation rates ($\gamma=1.2$ and $\gamma=1.5$~THz) leads to a larger range of field strengths for ferroelectricity.
These results indicate that efficient heat dissipation is necessary to observe THz field-induced ferroelectricity.
In the SM, we discuss how to estimate the dissipation rate $\gamma$ from $ab~initio$ molecular dynamics simulation~\cite{SM}.
    
Our results reproduce the experimental observation of THz field-induced ferroelectricity in SrTiO$_3$~\cite{Li2019}.
In the experiment, a single-cycle THz pulse was applied on SrTiO$_3$ and resulted in a non-oscillatory SHG signal at optical frequencies, signature of the ferroelectric phase.
Similar to this observation, we also obtained the non-oscillatory SHG signal and revealed the mechanism underlying this phenomenon.
In the same THz field-induced SHG traces and optical depolarization signals, the experiment also observed coherent oscillations, indicating the excitation of the AFD and FES modes in the ferroelectric state.
Our simulation found that applied THz field pulse excites the AFD mode, while its effect on the THz field-induced ferroelectricity is negligible at low intensity~\cite{SM}.
Considering the relation between the SHG signal and the FES mode coordinate ($\rm{SHG}(Q_f)=\rm{SHG}(-Q_f)$), the doubled frequency of the FES in the THz-induced SHG signal indicates a paraelectric state, which shows the motion of FES mode oscillation over the zero point ($Q_f=0$).
We propose that this behavior of frequency doubling can be experimentally observed under high temperature or high-intensity conditions, leading to a thermally excited paraelectric state rather than a ferroelectric state.
We further elaborate on the thermal effects such as thermal excitation of the AFD mode and a-axis lattice on the THz-field pulse-induced ferroelectricity in SrTiO$_3$ in the SM~\cite{SM,Li2019}.

In conclusion, we investigated THz field-induced ferroelectricity in quantum paraelectric SrTiO$_3$.
We solved the time-dependent 2D lattice Schr\"odinger-Langevin equation to simulate the action of a THz pulse on the material.
We found that a THz pulse with moderate field strength and considering a realistic dissipation rate induce ferroelectricity. 
The primary mechanism of this ferroelectricity lies in the excitation between ground and first excited states of the quantum paraelectric phase induced by the symmetry-breaking single-cycle THz pulse.
In agreement with previous experimental reports~\cite{Li2019,Nova2019}, we verified that the non-oscillatory SHG signal is a signature of ferroelectricity.
We also found that the resonant frequency of the THz pulse and realistic heat dissipation are required to trigger this non-equilibrium phase transition.
Our study reproduces the recent experimental observations and provides microscopic details that emphasize the importance of the quantum paraelectric phase in this phenomenon.

\begin{acknowledgments}
We are grateful for the discussions with Peter Littlewood, Antoine Georges, and Andrew Millis.
We acknowledge financial support from the European Research Council (ERC-2015-AdG-694097), Grupos Consolidados (IT1249-19), JSPS KAKENHI Grant Number 20K14382, the Cluster of Excellence 'CUI: Advanced Imaging of Matter' of the Deutsche Forschungsgemeinschaft (DFG) - EXC 2056 - project ID 390715994. 
D.S. and S.L. are supported by Alexander von Humboldt Foundation.
We also acknowledge support from the Max Planck–New York Center for Non-Equilibrium Quantum Phenomena.
The Flatiron Institute is a division of the Simons Foundation.
\end{acknowledgments}

\end{document}